\documentclass{article}

\newcount\commentcount \commentcount=0 
 \long\def\comment#1{\ifnum\commentcount=1 #1\fi}

\newcommand{\preprintsize}{
      \headheight=0pt                              
     \topmargin=0.2cm \headsep=0cm
      \oddsidemargin=0.2cm \evensidemargin=0.2cm  
      \textheight=22.5truecm \textwidth=16truecm      %
	  \setlength{\columnsep}{20pt}                  
}

\newtoks\reportnoregister \newtoks\eprintnoregister
\newcommand{\reportnumber}[1]{\reportnoregister={#1}}
\newcommand{\eprintnumber}[1]{\eprintnoregister={#1}}

\reportnumber{\mbox{}} 
\eprintnumber{\mbox{}} 

\newcommand{\reportid}{
   \begin{minipage}{13cm}\vspace{-3.0cm}  
     \begin{flushright}
      {\normalsize \the\reportnoregister \\[-.2cm]
	    \eprintstyle{\the\eprintnoregister}}\vspace{3.0cm} 
     \end{flushright}
   \end{minipage}\hspace{-13cm} }

\catcode`@=11   
\def\title#1{\gdef\@title{\reportid#1}}
\catcode`@=12   

\newcommand{\eprintstyle}[1]{\textsf{#1}} 

\newcommand{\journalfont}{\rm}  
\newcommand{\jou}[1]{{\journalfont #1\ }}
\newcommand{\joudef}[2]{\newcommand #1{\jou{\ignorespaces #2}}}

\joudef{\ajp}    { Am.~J.~Phys.}
\joudef{\aaa}    { Astron.\ Astrophys.}
\joudef{\aip}    { Adv.\ Phys.}
\joudef{\adm}    { Adv.\ Math.}
\joudef{\am}     { Ann.\ Math.}
\joudef{\apb}    { Ann.\ Phys.\ (Berlin)}
\joudef{\apny}   { Ann.\ Phys.\ (N.Y.)}
\joudef{\apj}    { Astrophys.\ J.}
\joudef{\apjs}   { Astrophys.\ J.\ Suppl.}
\joudef{\baps}   { Bull.~Am.~Phys.~Soc.}
\joudef{\cjp}    { Can.\ J.\ Phys.}
\joudef{\cmda}   { Celest.\ Mech.\ Dyn.\ Astron.}
\joudef{\cmp}    { Commun.\ Math.\ Phys.}
\joudef{\cqg}    { Class.\ Quantum Grav.}
\joudef{\faa}    { Funct.\ Anal.\ Appl.}
\joudef{\grg}    { Gen.\ Rel.\ Grav.}
\joudef{\ijmpd}  { Int.\ J.\ Mod.\ Phys.\ D}
\joudef{\ijtp}   { Int.\ J.\ Theor.\ Phys.}
\joudef{\invm}   { Invent.\ Math.}
\joudef{\jm}     { J.\ Math.}
\joudef{\jmp}    { J.\ Math.\ Phys.}
\joudef{\jpa}    { J.\ Phys.\ A}
\joudef{\jpamg}  { J.\ Phys.\ A:\ Math.\ Gen.}
\joudef{\jpdap}  { J.\ Phys.\ D:\ Appl.\ Phys.}
\joudef{\lrr}    { Living Rev. Relativity}
\joudef{\mnras}  { Mon.\ Not.\ R.\ Ast.\ Soc.}
\joudef{\mpla}   { Mod.\ Phys.\ Lett.\ A} 
\joudef{\nature} { Nature}
\joudef{\nc}     { Nuovo Cim.}
\joudef{\npb}    { Nuc.\ Phys.\ B}
\joudef{\ph}     { Physica}
\joudef{\pla}    { Phys.\ Lett.\ A}
\joudef{\plb}    { Phys.\ Lett.\ B}
\joudef{\pr}     { Phys.\ Rev.}
\joudef{\pra}    { Phys.\ Rev.\ A}
\joudef{\prb}    { Phys.\ Rev.\ B}
\joudef{\prc}    { Phys.\ Rev.\ C}
\joudef{\prd}    { Phys.\ Rev.\ D}
\joudef{\prep}   { Phys.\ Rep.}
\joudef{\prl}    { Phys.\ Rev.\ Lett.}
\joudef{\pnas}   { Proc.\ Natl.\ Acad.\ Sci.\ USA}
\joudef{\prsla}  { Proc.\ Roy.\ Soc.\ Lond.\ A}
\joudef{\ptp}    { Prog.\ Theor.\ Phys.}
\joudef{\ptps}   { Prog.\ Theor.\ Phys.\ Suppl.}
\joudef\rmp      { Rev.\ Mod.\ Phys.}
\joudef\spj      { Sov.\ Phys.\ JETP}
\joudef\jetpl    { JETP Lett.}

\newcommand{\gr}{04.20.-q}            
 
\newcommand{\einsteinmaxwell}{04.40.Nr}  
\newcommand{\grexperiments}{04.80.Cc}

\newcommand{\electrons}{14.60.Cd} 
\newcommand{\nucleons}{14.20.Dh}  

\catcode`@=11

\newcommand\eqalign[1]{\null\,\vcenter{\openup\jot\m@th
  \ialign{\strut\hfil$\displaystyle{##}$&$\displaystyle{{}##}$\hfil
      \crcr#1\crcr}}\,}
\newcommand\meqalign[1]{\null\,\vcenter{\openup\jot\m@th
  \ialign{\strut\hfil$\displaystyle{##}$&&$\displaystyle{{}##}$\hfil
      \crcr#1\crcr}}\,}
\def\ps@reportnumber{%
    \let\@oddfoot\@empty\let\@evenfoot\@empty
    \def\@oddhead{\hfil\rightmark}}
	
\catcode`@=12   

\newdimen\arrayruleHwidth
\makeatletter
\newcommand\Hline{\noalign{\ifnum0=`}\fi\hrule \@height \arrayruleHwidth
  \futurelet \@tempa\@xhline}
\makeatother

\newcommand\thickbaselines{\baselineskip=20pt\lineskip=3pt\lineskiplimit=3pt}

\catcode`@=11

\renewcommand\matrix[1]{\null\,\vcenter{\thickbaselines\m@th
    \ialign{\hfil$##$\hfil&&\quad\hfil$##$\hfil\crcr
      \mathstrut\crcr\noalign{\kern-\baselineskip}
      #1\crcr\mathstrut\crcr\noalign{\kern-\baselineskip}}}\,} 
\catcode`@=12   

\renewcommand{\d}{{\rm d}} 

 \newcommand\Fscr{{\cal F}}

\newcommand\Oscr{{\cal O}}

\newcommand\undersim[1]{\mathop{\vtop{\ialign{##\crcr
     $\hfil\displaystyle{#1}\hfil$\crcr\noalign
     {\kern1pt\nointerlineskip}\hbox{$\hfil\sim\hfil$}\crcr
     \noalign{\kern1pt}}}}}

\preprintsize

\usepackage{amsmath}
\usepackage{graphicx}
\usepackage{txfonts}

\newcommand{\gfactor}{{$g$-factor}}

\newcommand{\cm}{\text{cm}}
\newcommand{\fm}{\text{fm}}
\newcommand{\barn}{\text{b}} 

\newcommand{\eV}{\text{eV}}

\newcommand{\lspin}[1]{a_\text{#1}} 

\newcommand{\PhiG}{\Phi_\text{G}} 
\newcommand{\PhiE}{\Phi_\text{E}} 
\newcommand{\PhiC}{\Phi_\text{C}} 

\newcommand{\rstar}{\rho_0}

\begin{document}
\bibliographystyle{prsty}



\title{GRAVITATIONALLY INDUCED ELECTROMAGNETISM \\ AT THE COMPTON SCALE}
\vspace{20pt}
\author{Kjell Rosquist \\[20pt]
\emph{\normalsize Department of Physics, AlbaNova University Center} \\
\emph{\normalsize Stockholm University, 106 91 Stockholm, Sweden} \\
\eprintstyle{\normalsize kr@physto.se} \\[2cm]
%
\begin{minipage}[t]{0.8\linewidth}\small{
It is shown that Einstein gravity tends to modify the electric and magnetic fields appreciably at length scales of the order of the Compton wavelength. At that scale the gravitational field becomes spin dominated rather than mass dominated. The gravitational field couples to the electromagnetic field via the Einstein-Maxwell equations which in the simplest model (Kerr-Newman) causes the electrostatic field of charged spinning particles to acquire an oblate structure relative to the spin direction. For electrons and protons, the Coulomb field is therefore likely to be modified by general relativity at the Compton scale. In the Kerr-Newman model, the magnetic dipole is known to correspond to the Dirac \gfactor, $g=2$. Also, the electric dipole moment vanishes, in agreement with current experimental limits for the electron. Quantitatively, the classical Einstein-Maxwell field represented by the Kerr-Newman solution models the magnetic and electric dipoles of the electron to an accuracy of about one part in $10^{-3}$ or better taking into account also the anomalous magnetic moment. Going to the next multipole order, one finds for the Kerr-Newman model that the first non-vanishing higher multipole is the electric quadrupole moment which acquires the value ${-}124 \,\barn$ for the electron. Any non-zero value of the electric quadrupole moment for the electron or the proton would be a clear sign of curvature due to the implied violation of rotation invariance. There is also a possible spherical modification of the Coulomb force proportional to $r^{-4}$. However, the size of this effect is well below current experimental limits. The corrections to the hydrogen spectrum are expected to be small but possibly detectable. }
\end{minipage}}

\date{}

\maketitle

\centerline{\bigskip\noindent PACS numbers: \electrons, \nucleons, \gr,
            \einsteinmaxwell, \grexperiments }
\clearpage

\section{Introduction}

Gravitational forces are generally regarded as negligible for atomic and nuclear physics. This belief is based on the Newtonian argument (see e.g.\ Wheeler \cite{Wheeler:1962}) that the energy $GM^2/r$ of the gravitational interaction  between two protons is roughly $4 \times 10^{-30} \,\eV$ at the nucleon Compton wavelength.  This is extremely small compared to the other three forces. The weakest of these, the weak interaction, has an interaction energy at the same distance of approximately $10^4 \,\eV$.
It may be noted that it is the Newtonian form of the interaction energy which is used in the above argument. 
However, if we do not assume an \emph{a priori} knowledge of the validity of Newtonian gravity in this context we cannot be sure that the estimate is correct. 
In this note I will instead use a direct approach to estimate the gravitational and electromagnetic fields at short distances based on the Einstein-Maxwell field equations.
Primary properties of the electron which can be detected at long range are the mass ($M$), the charge ($Q$), the angular momentum per unit mass ($a = S/M$) and the magnetic dipole moment ($\mu$). The two latter quantities are related by the \gfactor, which for the electron has the value $g_\text{e} =2$ (the small QED corrections to $g_\text{e}$ are touched upon in the last section). It is well-known that the electrovacuum black hole solutions of the Einstein-Maxwell equations can be uniquely described by the Kerr-Newman metric \cite{Kerr:1963} \cite{Newman_etal:1965} \cite{Mazur:1982} \cite{Mazur:1984} (see \cite{Heusler:1998} for a review of the black hole uniqueness theorems). The Kerr-Newman solution is characterized by the three quantities $M$, $Q$ and $a$. In addition, it has a magnetic dipole moment corresponding to a \gfactor{} $g=2$, just as the electron. This fact was noted by Carter in 1968 \cite{Carter:1968}. An essential ingredient in the uniqueness theorems leading up to the Kerr-Newman geometry is that the solutions satisfy certain boundary conditions for black hole horizons.  The Kerr-Newman models are black holes only if they satisfy the inequality $M^2 \geq Q^2 + a^2$. By contrast, the Kerr-Newman solutions with $M^2 < Q^2 + a^2$ have a naked ring-like singularity in their central region. For spinning elementary particles like electrons and nucleons, unlike black holes, the mass is dominated by the spin, $a \gg M$ (numbers are given at the end of this section).

While the Kerr-Newman metric is the unique solution for black hole configurations, there is no corresponding uniqueness theorem which can be applied to elementary particles like the electron. Ideally, it would be possible to prove such a theorem if appropriate boundary conditions are imposed, although for reasons given below, some deviations from the Kerr-Newman geometry are expected, unlike the situation for black holes. 
As noted above, the Kerr-Newman metric has the correct \gfactor{} for the electron, $g=2$. Moreover, it has been shown to be the only asymptotically flat solution of the Einstein-Maxwell equations for which the geodesic and Klein-Gordon equations can be solved by separation of variables \cite{Dadhich&Turakulov:2002}. The Dirac equation is also known to be separable on a Kerr-Newman background \cite{Chandrasekhar:1983}. The upshot of all this is that although other solutions exist, the Kerr-Newman geometry is by far the simplest which can serve as a candidate model for the external Einstein-Maxwell field of the electron. Based on these arguments I will use the Kerr-Newman model as prototype to investigate the Einstein-Maxwell field at microscopic distances down to some radius $r_0>0$. The robustness of this assumption will be further discussed in the following sections. The nature of the ring singularity at $r=0$ will be left aside. The classical description of the gravitational field is expected to break down anyway at sufficiently small distances. Because of the quantum nature of the angular momentum of elementary particles it seems that the gravitational field itself will acquire some quantum aspect by being so closely tied to the spin. However, I will take the point of view here to pursue the classical non-quantum description. This will lead to conclusions which are possible to test experimentally.

It is also of interest to consider the Einstein-Maxwell field of the proton. However, there is a complication due to the fact that its \gfactor{} is $g_\text{p} =5.59$ and so is almost three times larger than that of the electron. This means that the Kerr-Newman geometry is not a good model for the proton's Einstein-Maxwell field. However, it should still be sufficient to use this model for the purpose of making rough estimates of the sizes of various effects. With this caveat in mind we will use the Kerr-Newman geometry as a model for the external Einstein-Maxwell field of the proton as well. The range of validity, specified by $r_0$, of this geometry is not an easy question but we can at least say that for the proton it would be shaky to consider the region inside the radius of the proton at about $1\,\fm = 10^{-13}\cm$. For the electron, the experimental situation becomes less clear for radii less than $10^{-18}\cm$, where it is not known if the electron can still be considered as a point particle. There is also the issue of vacuum polarization at the Compton radius and below. However, it is not our aim in this paper to propose a quantum description, but rather to obtain a better understanding of the classical underpinnings, such as for example the validity of the use of the Coulomb potential at short distances. However, since the picture which emerges differs substantially from the traditional one, it may indicate directions for modifying the quantum theory as well.

There are three characteristic scales which govern the gravitational and electromagnetic fields of the electron and the proton. Using geometric units (see e.g.\ \cite{Misner_etal:1973}), they are the mass radius\footnote{For macroscopic systems it is common to use $2M$ as the gravitational radius of an object. That is the value of $r$ which gives the location of the horizon of a Schwarzschild black hole with mass $M$. However, this is not relevant in the present context.} associated with $M$, the charge radius associated with $Q$ and the spin radius\footnote{This term is taken from \cite{Pekeris:1987}.} associated with $a$. The value of the charge radius is
\begin{displaymath}
   Q \rightarrow e = 1.38 \times 10^{-34} \,\text{cm} \ .
\end{displaymath}
For the electron and the proton, the other two lengths are
\begin{align*}
&\text{\textnormal{\emph{The electron mass radius}:}}&
M \rightarrow m_\text{e} &= 6.76 \times 10^{-56} \,\text{cm} \\[5pt]
&\text{\textnormal{\emph{The proton mass radius}:}}&
M \rightarrow m_\text{p} &= 1.24 \times 10^{-52} \,\text{cm} \\[5pt]
&\text{\textnormal{\emph{The electron spin radius}:}}&
a \rightarrow \lspin{e} &= \dfrac{\lambdabar_\text{e}}{2}
= \dfrac{\hbar}{2m_\text{e}} = 1.93 \times 10^{-11} \,\text{cm} \\[5pt]
&\text{\textnormal{\emph{The proton spin radius}:}}&
a \rightarrow \lspin{p} &= \dfrac{\lambdabar_\text{p}}{2} = \dfrac{\hbar}{2m_\text{p}} = 1.05 \times 10^{-14} \,\text{cm}
\end{align*}
where $\lambdabar$ is the (reduced) Compton wavelength. Note that $Ma = \hbar/2$ and that $2m_\text{e} a_\text{e} /e^2 = \alpha^{-1}$ where $\alpha$ is the fine structure constant. Note also that the three scales are hugely different; $M \ll Q \ll a$ with $Q/M \sim 10^{21}$ and $a/Q \sim 10^{23}$ for the electron.

\section{The gravitational and electromagnetic fields at atomic \\ and subatomic distances}

A prevailing view today is that classical (non-quantum) Einstein gravity can be trusted in the sub-Planckian regime $E \ll E_\text{Planck}$ corresponding to length scales $\ell \gg \ell_\text{Planck} = 1.6 \times 10^{-33} \cm$.\footnote{The Planck scale is defined by $\ell_\text{Planck} = \hbar^{1/2} = \alpha^{-1/2}e$. It is interesting to note the approximate equality (within a factor of ten) between the Planck scale and the unit of charge.} It is expected that the spacetime geometry in this sub-Planckian regime will be the arena of all non-gravitational physical processes. According to this view, gravitational interactions will only be important for microphysics at and below the Planck length where a quantum theory of gravity will be needed. In fact it turns out that in the simplest models, the gravitational field will deviate from its Newtonian form already at length scales of the order of the Compton wavelength. This will in turn lead to modifications of the electromagnetic field at the same distances. In particular the Coulomb form of the electrostatic interaction will break down at the Compton scale. This can be considered as a gravitationally induced electromagnetic effect. Such effects have been considered before in the context of gravitational radiation \cite{Johnston_etal:1973} \cite{Ruffini:1978}. To illustrate the effect we will estimate the gravitational and electric fields for the electron and the proton. For reasons of brevity, the magnetic field will not be discussed in this paper. The gravitational and electromagnetic fields should satisfy the Einstein-Maxwell field equations \cite{Misner_etal:1973}
\begin{equation}\label{field_equations}
  R_{\mu\nu} = \frac{\kappa}{4\pi} \left(F_{\mu\rho}F_\nu{}^\rho
     -\tfrac14 g_{\mu\nu}F_{\rho\sigma}F^{\rho\sigma} \right) \ ,
\end{equation}
which relate the Ricci curvature on the left hand side to the electromagnetic energy-momentum tensor on the right hand side. Here $\kappa$ is the Einstein gravitational constant which has the value $8\pi$ in geometric units.
These equations determine both the metric $g_{\mu\nu}$ and the electromagnetic field tensor $F_{\mu\nu}$.\footnote{Up to a duality rotation of the electromagnetic field \cite{Misner_etal:1973}.} Note first that a Coulomb force together with the Schwarzschild metric does not correspond to a solution of \eqref{field_equations}. In fact, if we assume spherical symmetry, then the only charged asymptotically flat solution of the Einstein-Maxwell equations is the Reissner-Nordstr\"om metric (see e.g.\ \cite{Misner_etal:1973}). It then follows that the gravitational potential switches from the mass dominated Newtonian $M/r$ form to the charge dominated form $Q^2/r^2$ at $r= Q^2/(2M)$. This can be seen from \eqref{PhiG} below by setting $a=0$. The switch from mass to charge domination happens at $r = e^2/(2m_\text{e}) = 1.4 \times 10^{-13} \cm$ for the electron (i.e.\ half the classical electron radius).  Furthermore, we know from studies of black hole physics for example, that the angular momentum of a source has profound effects on the gravitational field. In view of the fact that the spin of the electron and the proton dominates both the mass and the charge by huge factors, it seems to be an inescapable conclusion that it is the spin which will determine the gravitational field at the Compton scale. It is often argued that the spin is a quantum phenomenon which cannot be incorporated in a classical context. However, it is also well established that the spin couples to the orbital angular momentum in just the right way to conserve the total angular momentum. Here, we take the view to explore the possibility that the spin should be included in the classical Einstein-Maxwell equations to obtain the combined gravitational and electromagnetic fields of elementary particles at distances larger than $r>r_0$. The traditional view, to exclude the spin, is seemingly a much more drastic assumption given the fact that the spin dominates both the mass and the charge by the huge factors given above.

As discussed in the introduction, I assume that the gravitational field of an electron is given approximately by the Kerr-Newman solution (see \cite{Misner_etal:1973} \cite{Carter:1973} \cite{Cherubini_etal:2002} for details about its connection and curvature). Using Boyer-Lindquist coordinates \cite{Boyer&Lindquist:1967} we can then write the metric as
\begin{equation}
   g = -(L^0)^2  + (L^1)^2 + (L^2)^2 + (L^3)^2 \ ,
\end{equation}
where $L^\mu$ is a Lorentz co-frame given by
\begin{equation}\label{Lframe}
   L^0 = \frac{\sqrt{\Delta}}{\rho}(\d t-a\sin^2\!\theta\d\phi) \ ,\quad
   L^1 = \frac{\rho}{\sqrt{\Delta}} \d r \ ,\quad
   L^2 = \rho\d\theta \ ,\quad
   L^3 = \frac{\sin\theta}{\rho}(\rstar^2\d\phi-a\d t) \ ,
\end{equation}
where
\begin{equation}
   \Delta = r^2 - 2Mr + Q^2 + a^2 \ ,\qquad
   \rho^2 = r^2 + a^2 \cos^2\!\theta \ ,\qquad
    \rstar^2 = r^2 + a^2 \ .
\end{equation}
The electromagnetic field can be specified by the 4-potential which can be taken as \cite{Misner_etal:1973}
\begin{equation}
 A = -\frac{Qr}{\rho\sqrt{\Delta}} \, L^0
   = -\frac{Qr}{\rho^{2}}(\d t - a\sin^2\!\theta\d\phi) \ .
\end{equation}
The field itself is then given by the relation $F_{\mu\nu}\d x^\mu \wedge \d x^\nu = 2\d A$.

To calculate measurable quantities we must first make a choice of observer frame. It so happens that there is a preferred frame which corresponds to objects which are static with respect to the static observers at infinity (``distant stars"). The choice of observer frame is actually a rather subtle issue which we will return to when discussing the multipole expansion of the electric field in section \ref{electric}. It effectively requires that we define some flat background geometry (``laboratory frame") which serves as a reference frame. Our choice of the static observers is supported by the optical analogue formulation of the Schwarzschild geometry given in \cite{Rosquist:2004}. In these analogue models, the static observer frame coincides with the laboratory frame. We follow the standard procedure and declare an object in the Kerr-Newman spacetime to be a static object\footnote{Known as a static observer \cite{Misner_etal:1973} in a macroscopic context.} if its spatial coordinates $r, \theta, \phi$ are all constant. The 4-velocity of such a static object is given by
\begin{equation}
   u = f^{-1/2} \frac{\partial}{\partial t} \ ,
\end{equation}
where
\begin{equation}\label{PhiG}
   f = 1-2\PhiG \ ,\qquad  \PhiG
     = \frac{2Mr-Q^2}{2\rho^2} = \frac{2Mr-Q^2}{2(r^2+a^2\cos^2\!\theta)} \ ,
\end{equation}
and $\PhiG$ can be loosely regarded as a generalized gravitational potential. In the limit $r \rightarrow \infty$ it goes over into Newton's potential $\PhiG \rightarrow M/r$. It should be noted that $f>0$ for all values of the Boyer-Lindquist coordinates. This implies that there is no ergoregion \cite{Misner_etal:1973} in this geometry (at least not in the Boyer-Lindquist coordinate patch). Therefore the static observers are defined for all values of the Boyer-Lindquist coordinates. 

To calculate a vectorial physical quantity such as the electric field we should express it in a Lorentz frame adapted to the static observer field. Note that the co-frame $L^\mu$ in \eqref{Lframe} is boosted in the $\phi$-direction with respect to the static object frame. However, as discussed below, the electric field measured by a static observer has only components along the 1- and 2-directions which both lie in the rest frame of the static object. Therefore the two relevant frame vectors we need are
\begin{equation}\label{Lvectors}
   L_1 = \frac{\sqrt{\Delta}}{\rho} \frac{\partial}{\partial r} \ ,\qquad
   L_2 = \frac1{\rho} \frac{\partial}{\partial\theta} \ .
\end{equation}

\section{Multipole expansion of the electric field}\label{electric}

Performing a multipole expansion is an essential tool to understand the physical effects of gravitational and electromagnetic fields. Although such expansions are more or less straightforward in flat space, this is not the case in curved spacetimes (see for example Thorne \cite{Thorne:1980} and Simon \cite{Simon:1984}). In flat space, the multipole expansion is often defined in terms of the spherical harmonics which represent solutions of the Laplace equation. However, in curved space, the Laplace equation is only relevant in the limit $r \rightarrow \infty$. There are basically two issues which complicate the situation when curvature is present. The first is the choice of background or laboratory geometry. The second is the choice of coordinates. Actually, these issues are related since once a background geometry has been chosen, there is always the natural choice of spherical coordinates corresponding to any Cartesian coordinate system. This leaves only the Poincar\'e group as the remaining freedom. In practice, there is usually a natural choice of Cartesian coordinate frame in which the object under study is at rest. Choosing the origin as the center of mass further restricts the freedom as does an alignment along the axis of symmetry for axisymmetric systems. This leaves finally at most a subgroup of the rotation group as the remaining freedom. Some aspects of the physical significance of choosing an appropriate background have been discussed by  Penrose \cite{Penrose:1980} and by Gao and Wald \cite{Gao&Wald:2000}. Curiously, this issue has not been of much concern in works on multipole expansions, although Thorne \cite{Thorne:1980} remarked that his multipoles had other values than those of other workers. A common thread in all approaches appears to be the desire to express the expansion entirely in terms of spherical harmonics (see e.g.\ \cite{Simon:1984}). In my view, this is too restrictive since it automatically excludes, for example, spherical terms of the type $r^{-2}$ which a priori could have physical significance. From a physical point of view, there is no ambiguity in the choice of background, we should use the uniquely defined laboratory frame! The problem is to identify the laboratory frame in a given physical problem. Having done that there is still the issue of finding the relation between the spherical coordinate system of the laboratory and the coordinates used to specify the geometry of the physical system we are interested in. 

\begin{figure}[!t]\centering
  \begin{minipage}{0.8\textwidth}
      \hspace{-10mm}\mbox{\includegraphics[width=0.5\textwidth]{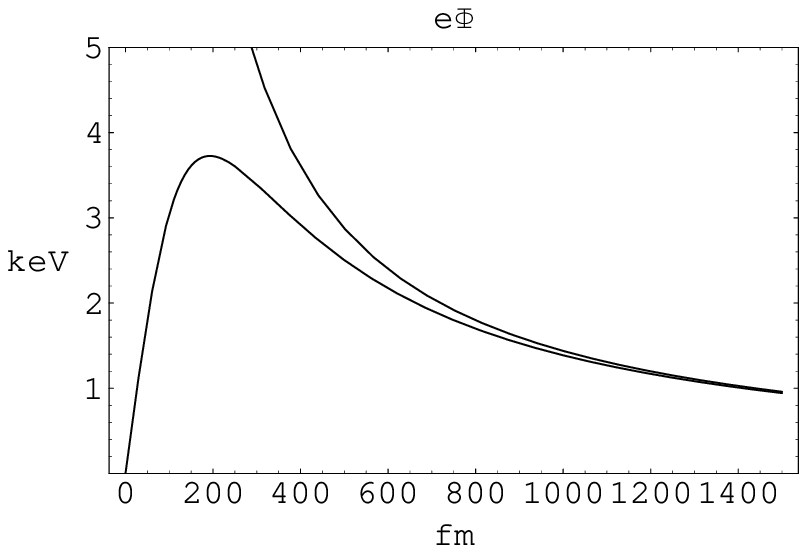}
            \hspace{5mm}
            \includegraphics[width=0.5\textwidth]{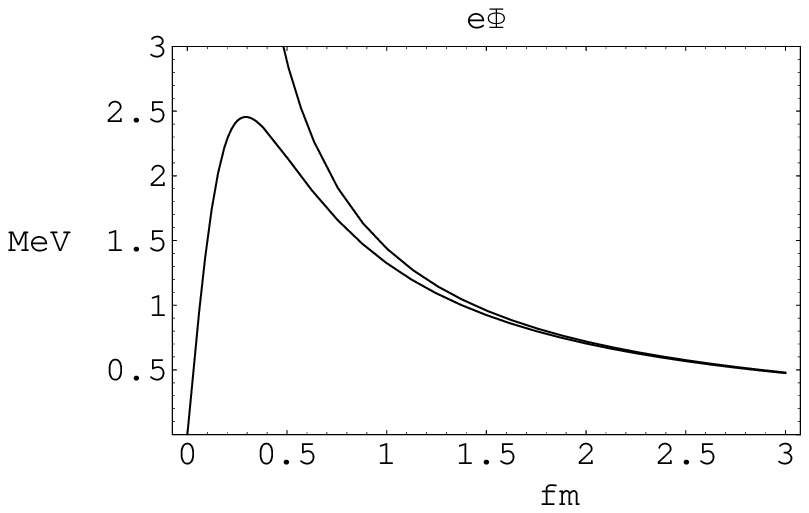}}
  \caption{\small
The Einstein electrostatic potential energy for a test particle with unit charge along the spin axis is plotted as a function of $r$ together with the Coulomb potential for comparison. The electron potential is shown in the left panel and the proton in the right. The latter should only be taken as a qualitative indication of the electric field of the proton. The curve for the proton has been computed with an adjusted spin radius $a \rightarrow (g_\text{p}/2) a$ to account for an expected enhancement of higher electromagnetic multipoles due to the \gfactor. The scale along the horizontal axis is in units of $1 \fm = 10^{-13} \cm$ (beware that the coordinate distance used here is not exactly equal to laboratory distance as discussed in the text). Note that the reduced Compton wavelength is $\lambdabar_\text{e} = 386\,\fm$ for the electron and $\lambdabar_\text{p} = 1.32 \,\fm$ for the proton. }
    \label{VE_axis}
  \end{minipage}
\end{figure}

\begin{figure}[!t]\centering
  \begin{minipage}{0.8\textwidth}\centering
      {\includegraphics[width=0.6\textwidth]{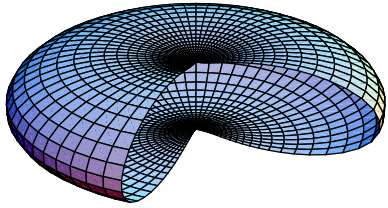}
       \vspace{-9mm}}
    \caption{\small
A spherical plot of the electrostatic potential at $r = a/2 = \lambdabar_\text{C}/4$  (for $g=2$). For each point on the surface, parametrized by the angles $(\theta,\phi)$, the value of the potential corresponds to the distance to the origin, $r=0$. Only values of the potential in the directions $0 \leq \phi \leq 3\pi/2$ are shown to make the extent of the pinching along the axis more evident.}
    \label{VE_spherical}
  \end{minipage}
\end{figure}

A multipole expansion of the electric field could start either from the electric field itself or from an electrostatic potential. Both these cases have their problems but we choose to use the potential approach as it seems somewhat easier to handle. It is not a straightforward procedure even to define an electrostatic potential in a curved spacetime. We start by considering the Lorentz force on a test charge $Q'$ with 4-velocity $u^\mu$. This force is given by the expression $\Fscr_\mu = Q' F_{\mu\nu} u^\nu$. The work needed to move the charge a distance $\d x^\mu$ then becomes
\begin{equation}
   \delta W = \Fscr_\mu \d x^\mu = Q'F_{\mu\nu}u^\nu\d x^\mu \ .
\end{equation}
To define a potential we must require that this work is a closed 1-form meaning that the condition
\begin{equation}\label{electrostatic_observers}
   \d (F_{\mu\nu} u^\nu \d x^\mu) = 0
\end{equation}
must be satisfied. This equation therefore defines observers for which an electrostatic potential can be defined. Turning now to the Kerr-Newman geometry and choosing the static observer frame as discussed above, a short calculation gives
\begin{equation}\label{deltaW_KN}
   \delta W/Q' = F_{\mu\nu} u^\nu \d x^\mu = u^t \d A_t
               = \d (u^t A_t) - A_t \d u^t \ .
\end{equation}
It can be verified by direct calculation that $\d (u^t \d A_t)$ does not vanish identically implying that the static observer frame is not associated with an electrostatic potential. However, in the limit $r \rightarrow \infty$ where $\d u^t \rightarrow 0$ we can still hope to use $\PhiE = -u^t A_t$ as potential. This hope can actually be substantiated by expansion of the two terms in \eqref{deltaW_KN} in powers of $1/r$. We therefore define the electrostatic potential by
\begin{equation}\label{PhiE}
   \PhiE = -u^t A_t = \frac{Qr}{f^{1/2}\rho^2}
        = \frac{Qr}
         {\sqrt{(r^2-2Mr+Q^2+a^2\cos^2\!\theta)(r^2+a^2\cos^2\!\theta)}} \ ,
\end{equation}
keeping in mind that it can only be used up to a certain expansion order to be specified later on. Note that $\PhiE$ approaches the Coulomb form at large distances. The expression \eqref{PhiE} is valid for objects which are non-relativistic with respect to the static objects. In practice this means non-relativistic in the laboratory since $f=1$ to extremely good accuracy for macroscopic distances. The form of the potential is illustrated in figures \ref{VE_axis} and \ref{VE_spherical}. When interpreting these figures, it should be kept in mind that the coordinate $r$ is different from the Euclidean radial variable used in the lab. However, at $r \sim a$ it is of the same order as the Euclidean radius so the diagrams should give a qualitatively correct picture down to the Compton scale. The precise relation between $r$ and measured values of e.g.\ the radius of the proton is a somewhat subtle issue related to the choice of observer frame.
Since the potential is a function of $\cos^2\!\theta$, there is no electric dipole moment. Recent measurements \cite{Regan_etal:2002} of the electric dipole moment of the electron have set an upper limit of $|d_\text{e}| \le 1.6 \times 10^{-27} e \,\cm$. From the general form of the moment structure (see \cite{Hoenselaers&Perjes:1990} and \cite{Sotiriou&Apostolatos:2004}) one would expect a non-zero dipole to have a size of the order $\sim ea$. It is therefore natural to express the upper limit in a dimensionless way as $|d_\text{e}|/(ea_\text{e}) \le 8.3 \times 10^{-17}$. Expansion in $1/r$ of the potential $\PhiE$ gives
\begin{equation}\label{PhiE_expansion}
   \PhiE/Q = \left[\frac1r - \tfrac23 a^2 P_2(\cos\theta) \frac1{r^3} \right]
            +\left[\frac{M}{r^2} + 
             \left(\tfrac32M^2 - \tfrac12Q^2 - \tfrac13a^2 \right)\frac1{r^3}
             \right] + \Oscr(r^{-4}) \ ,
\end{equation}
where $P_2$ is a Legendre polynomial. In \eqref{PhiE_expansion} we have grouped together terms corresponding to a standard multipole expansion between the square brackets to the left. The remaining terms between the square brackets to the right are due the fact that the potential is not a solution of the Laplace equation as expected in a curved geometry. Comparison with the expansion of the term $u^t \d A_t$ in \eqref{deltaW_KN} shows that the omission of that term does not affect the angle dependent terms in \eqref{PhiE_expansion} at the quadrupole level.  As for the spherical terms, their structure is not affected and only the numerical values of the $Q$-dependent coefficients do get changed for terms of the order $r^{-2}$ and below. As discussed below these terms are dominated by $a$-dependent terms at this level. Therefore the expansion in \eqref{deltaW_KN} is reliable up to the specified order ($r^{-3}$).

It has already been noted that the exact form of the expansion \eqref{PhiE_expansion} depends on the choice of observer frame.  In addition to the choice of frame, there is the problem of interpretation of the coordinates in relation to the spacetime of the laboratory. In particular, a preliminary investigation indicates that the spherical coordinate radius in the static observer frame is given by $\hat r = (r^2+ a^2\sin^2\!\theta )^{1/2}$. 
Performing the multipole expansion in $\hat r$ shows that the multipoles in \eqref{PhiE_expansion} will remain unchanged. The second term in \eqref{PhiE_expansion} corresponds to an electric quadrupole moment given by $q = -\tfrac13 a^2$. For the electron, this gives $q_e = -124 \,\barn$ ($1 \barn = 10^{-24} \cm^2$) in conventional units. This is actually quite large by the standards of nuclear electric quadrupole moments. For example, the measured value for the deuteron is $q_\text{D} = 0.00286 \,\barn$ \cite{Firestone:1996} which is smaller than the predicted electron value by a factor of about $10^{-5}$.

The fact that $q<0$ indicates that the electric potential corresponds to an oblate charge distribution (cf.\ figure \ref{VE_spherical}). According to the Wigner-Eckart theorem, a spin $\frac12$ particle cannot have an electric quadrupole moment (see e.g.\ \cite{Merzbacher:1961}). However, that theorem is a statement about eigenstates of the angular momentum operator in flat space. Put another way, a measurement of a non-zero electric quadrupole moment for the electron or the proton would be a clear signature of spacetime curvature.
For the proton one would expect a value for the quadrupole about $(g_\text{p}/2)^2 (a_\text{p}/a_\text{e})^2 \sim 10^{-6}$ times the electron value. It is interesting that the deuteron quadrupole, although prolate, has a size which is not very far off the expected value for the proton. Apparently, no measurements of the quadrupoles of the electron or the proton have been attempted.

Another effect of the curved geometry is that the potential $\PhiE$ fails to be a solution of the Laplace equation. This is manifested in \eqref{PhiE_expansion} by the appearence of higher order terms which do not correspond to a standard multipole expansion. The first two of these are the second order term $M/r^2$ and the third order term $-a^2/(3r^3)$ (neglecting $Q$ and $M$). As can readily be checked, the third order term dominates the second order one even at macroscopic distances. The result is that the Coulomb potential acquires a modification of the form
\begin{equation}\label{Coulomb_modification}
   \frac{Q}r \rightarrow \frac{Q}r\left(1 - \frac{a^2}{3r^2}\right) \ .
\end{equation}
Experimenters often model deviations from Coulomb's law by assuming a potential of the form $r^{-(1+s)}$ where $s$ is the parameter to be determined experimentally \cite{Tu&Luo:2004}. This means that for a given candidate potential $\Phi$, the parameter $s$ can be expressed as
\begin{equation}
   s = -r \frac{\d}{\d r} \ln \left( \frac{\Phi}{\PhiC} \right) \ ,
\end{equation}
where $\PhiC$ is the Coulomb potential. Now using the form \eqref{Coulomb_modification} we find to lowest order in $a^2/r^2$
\begin{equation}
   s = \frac{2a^2}{3r^2} \ .
\end{equation}
Setting $a=a_\text{e}$ and assuming a laboratory distance of $50\,\cm$ we then get a value for the correction of the order $s \sim 10^{-25}$. This is far below the present best limit $6 \times 10^{-17}$ \cite{Tu&Luo:2004}. 

To estimate the effect on the hydrogen spectrum we note that the correction to the potential energy from \eqref{PhiE_expansion} is of the order $\Delta V/V \sim (g_\text{p}/2)^2 (a/r)^2$. This leads to a change in the potential energy of the electron in the electric field of the proton of the order $\Delta V/V = (g_\text{p}/2)^2 (a_\text{p}/ l_\text{B})^2 \sim 10^{-13}$ where $l_\text{B}$ is the Bohr radius. The size of this correction is at the limit of present measurements \cite{Mohr&Taylor:2000}. There will also be a correction due to a change in the potential energy of the proton in the electron's electric field.  In addition, corrections to the hyperfine splitting also arise from changes in the magnetic field. The latter two effects are quite delicate to estimate since it would involve a recalculation of the Lamb shift using the  Kerr-Newman electromagnetic fields. Suffice it to say here that the suppression of the electric field in the spin direction is qualitatively consistent with the  QED induced softening of the Coulomb field a the Compton scale. It is also of interest to consider effects on the positronium spectrum. It turns out that the corrections are again approximately at the limit of the present experimental accuracy.

Turning now to the electric field itself, we find that for the static object, it is given by
\begin{equation}
   E_\mu = F_{\mu\nu} \,u^\nu = F_{\mu t} u^t \ ,
\end{equation}
leading to
\begin{equation}
   E_\mu \d x^\mu = \frac{Q(r^2-a^2\cos^2\!\theta)}
            {\rho^3\Sigma^{1/2}} \d r
     - \frac{Qa^2r\sin2\theta}
            {\rho^3\Sigma^{1/2}} \d\theta \ ,
\end{equation}
where we have defined
\begin{equation}
   \Sigma = r^2 - 2Mr + Q^2 + a^2\cos^2\!\theta \ .
\end{equation}
Expressed in the static object rest frame \eqref{Lvectors} the electric field vector can be written
\begin{equation}
   E = E^1 L_1 + E^2 L_2 = \frac{Q\Delta^{1/2}(r^2-a^2\cos^2\!\theta)}
            {\rho^4\Sigma^{1/2}}\,L_1
     - \frac{Qa^2r\sin2\theta}{\rho^4\Sigma^{1/2}}\,L_2\ .
\end{equation}
Let us now consider the radial electric force per unit mass on a test particle with mass $M'$ and charge $Q'$
\begin{equation}
   \Fscr_\text{E}^1(r,\theta) = \frac{Q'}{M'}E^1
    = \frac{Q'Q\Delta^{1/2}(r^2-a^2\cos^2\!\theta)}
            {M'\rho^4\Sigma^{1/2}} \ .
\end{equation}
At infinity this reduces to $\Fscr_\text{E}^1 = (Q'/M')(Q/r^2)$ which is just Coulomb's force (per unit mass). In the small $r$ limit, the behavior is $\Fscr_\text{E}^1(r,\pi/2) = (Q'/M')(a/r^{2})$ in the equatorial plane, while along the symmetry axis we have $\Fscr_\text{E}^1(0,0) = -(Q'/M')(Q/a^2)$. 

It is of interest to compare the radial electromagnetic and gravitational forces on a test particle in the small $r$ limit. The ratios between the forces is given by $(\Fscr_\text{E}/\Fscr_\text{G})_{\theta=\pi/2} = Q'/M'$ and $(\Fscr_\text{E}/\Fscr_\text{G})_{\theta=0} = -(Q'/M')(Q/M)$. Using $M=M'=m_\text{e}$ and $Q=Q'=e$ we find that $\Fscr_\text{E}/\Fscr_\text{G} \sim 10^{21}$ in the equatorial plane and $\Fscr_\text{E}/\Fscr_\text{G} \sim 10^{42}$ along the axis. Here the gravitational force $\Fscr_\text{G}$ on a static object has been computed as the mass times the acceleration. This shows that the electromagnetic forces dominate by huge factors, although the ratio in the equatorial plane is smaller than the Newtonian estimate $10^{42}$. However, the reason that the forces differ in strength is really due to the charge-to-mass ratio, not because the strengths of the gravitational and electromagnetic fields themselves are drastically different. On the contrary, if one compares the gravitational force per unit mass $\Fscr_\text{G}$ with the electric force per unit charge $(M'/Q') \Fscr_\text{E}$ one finds that they are equal along the axis. This means that the gravitational and electromagnetic fields themselves are of the same order. The gravitational force, however, is so much smaller than the electromagnetic force because of the exceedingly small factor $m/e$ for the elementary particles. This explains why the apparently weak gravitational field can give rise to sizable electromagnetic effects via the Einstein-Maxwell field equations. As a note of caution, it is important to bear in mind  that these estimates are valid only for test part particles. Considering, for example, the interaction of two Kerr-Newman particles is something very different, requiring a more sophisticated analysis.

\section{Discussion}

The identification of the Kerr-Newman geometry with the gravitational field of the electron appears to be the simplest and most natural assumption. But what about the quantum nature of the spin? Isn't it unreasonable to use a quantized property in the metric as though it were classical? Here, one could turn the table around and ask the same question about the quantized charge and for that matter about the quantized mass. Or, why should the spin (or the magnetic moment) be treated differently than the mass and the charge with respect to gravity in the microscopic domain? And if this is really so, in what situations should we ignore the spin and when should we use the macroscopic recipe for setting up the gravitational field? Another possible counter-argument is that the spin is vectorial while the charge and the mass are scalar quantities. However, this would still leave the question open why this fact should motivate a different treatment in the microscopic domain.

From a more quantitative point of view one could argue as follows. Suppose that the Kerr-Newman geometry is at least a good approximation at large distances for the electron's Einstein-Maxwell field. Then certainly, this approximation must break down at some distance, $r_0$, if for no other reason, because the metric has a curvature singularity at $\theta = \pi/2$, $r=0$. Now, the only free parameters in the Kerr-Newman metric are the mass, $m_\text{e}$, the charge, $e$, and the spin $m_\text{e} a_\text{e}$.
Moreover, we do know that the Kerr-Newman metric gives an accurate description of both the magnetic and the electric dipoles. Let us say we model the deviation from Kerr-Newman in the far field by adding an effective stress tensor, $T_{\mu\nu}^\text{eff}$, in the field equations. This tensor could emanate from QED corrections or quantum gravity for example.  Then we know that we can neglect $T_{\mu\nu}^\text{eff}$ at the dipole level. As discussed above the electric dipole has been constrained by experiments (\cite{Regan_etal:2002}) to be zero at an accuracy level of $\lesssim 10^{-16}$. Because of the methodology used in those experiments, they constrain only odd (parity violating) electric multipoles. The magnetic dipole on the other hand is measured to be (\cite{Mohr&Taylor:2000}) $\mu_\text{e}/(e a_\text{e}) = 1 + \alpha/(2\pi) + \Oscr(\alpha^{-2})$. The deviation from Kerr-Newman should therefore be of the order $\alpha/(2\pi) \sim 10^{-3}$.
However, the situation is complicated by the possible influence of higher multipoles which could in principle be affecting the cyclotron type experiments for measuring the electron's anomalous magnetic moment (see \cite{Kessler:1976} \cite{Van_Dyck_etal:1986} \cite{Van_Dyck:1990} for a description of how those experiments are made). The actual deviation from Kerr-Newman could then be different from that indicated above. We may conclude from this discussion that the influence of $T_{\mu\nu}^\text{eff}$ on the dipoles is probably at most $10^{-3}$ and maybe smaller. It then seems likely that the influence of $T_{\mu\nu}^\text{eff}$ on the next multipole, namely the electric quadrupole, is also small. Thus, even though the numerical value for the quadrupole would be affected at some level of approximation, the first few digits could well coincide with the Kerr-Newman values.

A main conclusion of this work is that the gravitational and electromagnetic fields may have an appreciable interaction at the Compton scale. I have shown that this leads to observable consequences. In particular, the most obvious test of these ideas is to measure the electric quadrupole moment of the electron. The impressive experimental limit set on the electric dipole moment suggests that it would not be beyond present techniques to perform such a measurement. The accurate values obtained for nuclear electric quadrupoles are also encouraging in this regard. In fact, the Kerr-Newman value of the electron's quadrupole is so large that it could well be hidden in existing experimental data. It would also be of interest to measure or set limits on the electric quadrupole moment of the proton. Any non-zero value of the electric quadrupole moment for the electron or the proton would signal the presence of curvature because of the violation of the Wigner-Eckart theorem and hence the implied breaking of rotational invariance.

The predicted value for the electric quadrupole moment of the electron is not a certain consequence of the Einstein-Maxwell equations. Other solutions which are more general than Kerr-Newman do exist (see e.g.\ \cite{Manko_etal:2000}). However, it would require a substantial amount of fine-tuning to make the quadrupole exactly zero. In principle, a vanishing quadrupole could be the result of some as yet unknown selection rule. For the proton though, being a composite particle, such fine-tuning would seem even more artificial.
Even if it is true that the Kerr-Newman metric is a good model for the Einstein-Maxwell field of the electron, it would be nice to have some more hard mathematical theorem to justify this assumption, for example along the lines of the uniqueness theorems for black holes. In the absence of such a theorem, it would still seem likely, given the assumption of spin domination, that the Kerr-Newman metric at the very least gives a good qualitative description of the gravitational field of the electron at distances down to nuclear length scales.

There appears to be only a few options, depending on the outcome of future experiments or reexamination of presently available experimental data. We could accept the lesson from general relativity and use the modified forms of the electric and magnetic fields in atomic and subatomic physics, or we could abandon or at least modify Einstein gravity at small distances, starting at some length scale above the Compton wavelength. In the latter case, a microscopic spin would have a different status vis-\`a-vis gravity than a macroscopic angular momentum.  A third but less likely alternative would be that even if we accept general relativity, nature somehow conspires to keep spacetime almost but not quite flat at microscopic scales.

It goes without saying that if gravity is really spin dominated at the Compton scale, then this would affect many aspects of physics. The questions which pop up in such a scenario are too many to be mentioned here. To name only a few, is the drastic changes of the electric and gravitational fields in the spin direction at the Compton wavelength scale an indication that the nuclear forces are more directly connected with the electromagnetic and gravitational forces than has been thought previously? This point is the subject of a paper in preparation where it will be suggested that the modified electromagnetic field can lead to bound states of two or more particles. In that picture, the nuclear binding energy scale will correspond to a minimum of roughly the same size as the 2 MeV peak in the proton potential (see figure \ref{VE_axis}). Another related issue is what happens to quantization if you can no longer assume a flat background. For example, in a consistent quantum theory, the electron creation operator should create its own patch of spacetime together with the other properties of the electron. Is quantum gravity entering through Compton's back door?

\vspace{15pt}
\noindent\emph{Endnote:}\newline
A scan of the literature reveals that several authors have considered the Kerr-Newman metric as a model for the electron's gravitational and electromagnetic fields. In particular, the late Chaim L.\ Pekeris noted the influence of the spin on the gravitational field at the Compton wave length \cite{Pekeris:1987}. Following that work Pekeris and Frankowski \cite{Pekeris&Frankowski:1989} tried to treat the electron as a Kerr-Newman geometry in its own right without an interior source. Using Chandrasekhar's separation \cite{Chandrasekhar:1983} of the Dirac equation on the Kerr-Newman background they studied the solutions but found that the states were unstable although in good agreement numerically with standard theory including the hyperfine levels (apart from an unexplained factor of two). Earlier, the role of the magnetic dipole in the gravitational field of the electron was emphasized by Martin and Pritchett \cite{Martin&Pritchett:1968}. Their approach was perturbative, with the gravitational constant $\kappa$ as perturbation parameter, but they took Schwarzschild rather than Kerr as the zeroth order metric. Redoing their analysis with the Kerr metric at zero order should lead to more interesting and physically relevant results.  Other authors have tried to construct a classical model for the electron with the Kerr-Newman metric as the exterior gravitational field glued to an interior extended charged rotating source (see e.g.\ \cite{Gron:1985} \cite{Ledvinka_etal:1999} and references therein).  In a recent paper, Arcos and Pereira \cite{Arcos&Pereira:2004} discuss implications of the topological structure of the Kerr-Newman geometry if taken as a model for the electron. Using a supersymmetric extension of the Wheeler-Feynman action-at-a-distance approach, Tugai and Zheltukhin \cite{Tugai&Zheltukhin:1995} found a prolate (rather than oblate as in Kerr-Newman) spin-dependent modification of the Coulomb field at the Compton scale.

\section*{Acknowledgements}

Many people have contributed in one way or another to the ideas which have led to this paper. I would like to mention in particular Lars Bergstr\"om and Remo Ruffini without whom this work would probably never have been started. Many thanks go to Lars Samuelsson for helpful and stimulating discussions during the final stages of preparation. I am also grateful to the members of the CoPS group for useful and enlightening debates.

\bibliography{kr}
\end{document}